\documentstyle[preprint,pra,aps]{revtex}

\begin{document}

\bibliographystyle{prsty}
\tighten

\title{Finite resolution measurement of the non-classical polarization
statistics of entangled photon pairs}

\author{Holger F. Hofmann}
\address{Department of Physics, Faculty of Science, University of Tokyo\\
7-3-1 Hongo, Bunkyo-ku, Tokyo113-0033, Japan}


\date{\today}

\maketitle

\begin{abstract}
By limiting the resolution of quantum measurements, the measurement
induced changes of the quantum state can be reduced, permitting  
subsequent measurements of variables that do not commute with the
initially measured property. It is then possible to experimentally
determine correlations between non-commuting variables.
The application of this method to the polarization statistics of entangled
photon pairs reveals that negative conditional probabilities between 
non-orthogonal polarization components are responsible for the violation
of Bell's inequalities. Such negative probabilities can also be observed
in finite resolution measurements of the polarization of a single photon. 
The violation of Bell's inequalities therefore originates from local 
properties of the quantum statistics of single photon polarization.
\end{abstract}


\section{Introduction}
Perhaps the most convincing proof of the non-classical nature of quantum
statistics is the violation of Bell's inequalities by a pair of entangled 
spin-1/2 particles \cite{Bel64}. Several experimental tests of these 
inequalities have been performed on pairs of entangled photons generated
either by two photon emission \cite{Fre72,Asp82} or by parametric 
down conversion \cite{Ou88,Shi88}. These experimental tests compare the 
polarization correlations 
of photon pairs emitted at the same time for different sets of orthogonal
polarizations. While no information about the relationship between 
non-orthogonal polarization directions of the single photon are revealed
in such measurements, the statistics obtained correspond to the quantum
theoretical prediction. Since the quantum formalism from which the violation
of Bell's inequalities is derived is widely accepted, one might wonder whether
it should not be possible to obtain a clearer understanding of the origin
of this non-classical effect by investigating the unique statistical 
connection between non-commuting quantum variables in more detail. 
In particular, finite resolution measurements can provide quantitative 
information about a quantum variable without destroying the quantum coherence
between different eigenstate components of that variable \cite{Hof00a}. 
By applying finite resolution measurements, it is therefore possible to
identify the non-classical correlations between non-commuting variables
directly \cite{Hof00b,Hof00c}. In the following, an experiment is proposed to 
determine the correlations between non-orthogonal polarizations
of entangled photon pairs. It is shown that the violation of Bell's 
inequalities results from the negative joint probabilities arising from
local non-classical correlations of the photon polarization. It is then 
possible to give a local interpretation of entanglement based on  
standard quantum mechanics. 

The rest of the paper is organized as follows.
In section \ref{sec:postulate}, the application of finite resolution 
measurements to the polarization of a single photon is discussed and
fundamental non-classical correlations are derived. In section 
\ref{sec:entangle}, the experimental setup for a measurement of entangled
photon pairs is presented and the statistical results of such a measurement
are derived. In section \ref{sec:quantcorr}, the non-classical features of
the statistics are identified and the implications concerning the nature
of entanglement are discussed. Finally, the conclusions are summarized in
section \ref{sec:concl}.

\section{Finite resolution measurements}
\label{sec:postulate}
\subsection{The generalized measurement postulate}
A measurement assigns a quantity to a system property based on
the observable action of the system on some measurement device. The 
uncertainty principle of quantum mechanics requires that this interaction 
between the system and the measurement setup introduces noise into
properties that do not commute with the measured variable. 
Therefore, the classical ideal of a complete determination of all
system properties is unattainable. Nevertheless it is possible to
obtain quantitative information on the correlations between non-commuting
variables by limiting the measurement resolution. 
Such a finite resolution measurement is described by the generalized 
measurement operator $\hat{P}_{\delta\!s} (s_m)$, which assigns a continuous 
measurement value $s_m$ to the operator $\hat{s}$\cite{Hof00a}. It reads
\begin{equation}
\hat{P}_{\delta\!s} (s_m) = \left(2\pi \delta\!s^2 \right)^{-\frac{1}{4}}
\exp\left(- \frac{(s_m-\hat{s})^2}{4\delta\!s^2}\right).
\end{equation} 
For a given initial state $\mid \psi_{\mbox{in}}\rangle$, 
the probability  $P(s_m)$ of obtaining a measurement result $s_m$ and the
state $\mid \psi_{\mbox{out}}\rangle$ after the measurement are then
given by
\begin{eqnarray}
P(s_m) &=& \langle \psi_{\mbox{in}} \mid \hat{P}^2_{\delta\!s} (s_m)
           \mid \psi_{\mbox{in}} \rangle \nonumber \\
\mid \psi_{out}\rangle &=& \frac{1}{\sqrt{P(s_m)}}
\hat{P}_{\delta\!s} (s_m) \mid \psi_{in}\rangle.
\end{eqnarray}
Note that this generalized measurement postulate does not restrict the
values of an operator variable to the eigenvalues of that operator. 
Eigenvalues emerge only in infinitely precise measurements. One of the 
fundamental problems in the discussion of quantum mechanics is that 
eigenvalues are often identified with ``elements of reality''\cite{EPR,Mer} 
regardless of the measurement context discussed. By assigning a continuous
measurement value to the operator variable in a finite resolution measurement
this identification is avoided, allowing a determination of quantitative 
results beyond the spectrum of its eigenvalues. 

\subsection{Finite resolution measurement of photon polarization}
The polarization of a single photon can be described in terms of the 
Stokes parameters $\hat{s}_i$. In terms of the circular polarization
eigenstates $\mid R \rangle$ and $\mid L \rangle$, the operators of the
three single photon Stokes parameters may be written as
\begin{eqnarray}
\hat{s}_1 &=& \hspace{0.2cm}
\mid L \rangle\langle R \mid + \hspace{0.2cm}\mid R \rangle\langle L \mid 
\nonumber \\
\hat{s}_2 &=& i \mid L \rangle\langle R \mid - i \mid R \rangle\langle L \mid
\nonumber \\
\hat{s}_3 &=& \hspace{0.2cm}
\mid R \rangle\langle R \mid - \hspace{0.2cm} \mid L \rangle\langle L \mid.
\end{eqnarray}
$\hat{s}_1$ represents the intensity difference between the x and y 
polarizations, $\hat{s}_2$ represents the intensity difference between the
polarizations along the diagonals between the x and y axes, and 
$\hat{s}_3$ represents the intensity difference between the circular 
polarizations. Since only one photon is considered, the eigenvalues of 
each Stokes parameter are $\pm 1$. 

A finite resolution measurement of photon polarization can be realized by 
using a polarization sensitive beam displacer that shifts the x-polarization
component relative to the y-polarization component of the light field.
The displacement of the photon trajectory in the beam displacer can be
interpreted as the action of the one-photon Stokes parameter $\hat{s}_1$. 
The polarization of the photon is then described by a 
continuous value $s_{1m}$ of the Stokes parameter $\hat{s}_1$ obtained from 
the measurement of the transversal photon position after the beam displacer.
The measurement resolution depends on the ratio of the displacement
and the width of the input beam. If the transversal profile of the
light field is Gaussian, the generalized measurement
postulate describes the single photon polarization statistics obtained
by measuring the polarization dependent displacement of the photon.
In terms of the circular polarization eigenstates, the measurement operator 
is given by
\begin{eqnarray}
\lefteqn{\hat{P}_{\delta\!s}(s_{1m}) = 
(2\pi\delta\!s^2)^{-\frac{1}{4}} 
\exp\left(-\frac{s_{1m}^2+1}{4\delta\!s^2}\right)}
\nonumber \\ &\times\big(& 
\;\cosh\left(\frac{s_{1m}}{2\delta\!s^2}\right) 
\left(\mid R \rangle \langle R \mid + \mid L \rangle \langle L \mid\right)
\nonumber \\ &&
+ \sinh\left(\frac{s_{1m}}{2\delta\!s^2}\right) 
\left(\mid R \rangle \langle L \mid + \mid L \rangle \langle R \mid\right)
\big).
\end{eqnarray}
This operator describes the changes in the quantum state of the single 
photon polarization conditioned by the finite resolution measurement of
the Stokes parameter $\hat{s}_1$.

\subsection{Joint measurements of non-orthogonal polarizations}
In order to measure the correlated non-orthogonal polarization components 
of a single photon, the finite resolution measurement can be combined with
a fully resolved polarization measurement. By rotating the polarization 
by an angle of $\pi/4$ and separating the x and y components as shown in 
figure \ref{setup1}, the eigenvalues of the Stokes parameter $\hat{s}_2$ are
measured. Two spatial patterns emerge, corresponding to the 
conditional distributions of continuous measurement results $s_{1m}$
of the Stokes parameter $\hat{s}_1$ 
associated with a final measurement of the eigenvalues $+1$ or $-1$ of 
the Stokes parameter $\hat{s}_{2}$.
The positive valued operator measure (POM) describing the joint measurement
of $\hat{s}_1$ and $\hat{s}_2$ is defined by projections onto the states
\begin{equation}
\mid s_{1m};s_2=\pm 1\rangle = 
\hat{P}_{\delta\!s}(s_{1m})
\frac{1}{\sqrt{2}}\left(\mid R \rangle \pm i \mid L \rangle\right).
\end{equation}
The joint probabilities $P(s_{1m};s_2=\pm1)$ for measuring a finite 
resolution value of $s_{1m}$ for the Stokes parameter $\hat{s}_1$ 
followed by an eigenvalue of $s_2=\pm 1$ for the Stokes parameter 
$\hat{s}_2$ is then given by
\begin{eqnarray}
\label{eq:joint}
P(s_{1m};s_2=\pm1) &=& 
|\langle s_{1m};s_2=\pm 1 \mid \psi_{\mbox{in}} \rangle|^2
\nonumber \\
&=& \frac{1}{2}
\left|\langle R \mid \hat{P}_{\delta\!s}(s_{1m})\mid\psi_{\mbox{in}} \rangle
\mp i \langle L \mid \hat{P}_{\delta\!s}(s_{1m})\mid\psi_{\mbox{in}} \rangle
\right|^2
, 
\end{eqnarray}
where $\mid \psi_{\mbox{in}}\rangle$ is an arbitrary initial state.
This POM thus assigns quantitative results to both Stokes parameters, allowing
a derivation of correlations between the polarization components of a single
photon.

If the light field entering the measurement setup shown in
figure \ref{setup1} is polarized along the diagonal between the x and y
axes, the initial photon state is given by
\begin{equation}
\label{eq:input}
\mid \psi_{\mbox{in}}\rangle = \frac{1}{\sqrt{2}}\left(\mid R \rangle 
                             + i \mid L \rangle \right).
\end{equation}
The joint probabilities of the measurement results $s_{1m}$ and $s_2$ can
then be determined using equation (\ref{eq:joint}). 
In its most compact form, it reads
\begin{eqnarray}
\label{eq:condprop}
P(s_{1m};s_2=+1) &=& \left(2\pi \delta\!s^2\right)^{-\frac{1}{2}}
                   \exp\left(-\frac{s_{1m}^2+1}{2\delta\!s^2}\right) 
                   \cosh^2\left(\frac{s_{1m}}{2 \delta\!s^2}\right)
\nonumber \\
P(s_{1m};s_2=-1) &=& \left(2\pi \delta\!s^2\right)^{-\frac{1}{2}}
                   \exp\left(-\frac{s_{1m}^2+1}{2\delta\!s^2}\right) 
                   \sinh^2\left(\frac{s_{1m}}{2 \delta\!s^2}\right).
\end{eqnarray}
Note that $P(s_{1m}=0;s_2=-1)$ is always exactly equal to zero, even if 
$\delta\!s$ is larger than one. Obviously, this result
is too exact to be explained in terms of a random measurement error
superimposed on classical statistics. The result for a measurement resolution 
of $\delta\!s=0.6$ is illustrated in figure \ref{condprop1}. The
peaks in $P(s_{1m};s_2=-1)$ are shifted to values of about $\pm 1.1$ and
the asymmetry of the peaks seems to favor even higher values. These results
can hardly be explained by statistics originating only from the eigenvalues
of $s_1=\pm1$. 

\subsection{Negative conditional probabilities and non-classical 
correlations in the polarization of single photons}
The non-classical features of the joint probabilities $P(s_{1m};s_2=\pm1)$
can be analyzed by expressing the result 
as a sum of shifted normalized Gaussian distributions 
\begin{equation}
G_{\delta\!s}(s_{1m}-d) := \left(2\pi \delta\!s^2\right)^{-\frac{1}{2}}
\exp\left(-\frac{(s_{1m}^2-d)^2}{2\delta\! s^2}\right). 
\end{equation}
In terms of these Gaussians, the joint probabilities read
\begin{eqnarray}
\label{eq:sumcp}
P(s_{1m};s_2=+1) &=& \frac{1}{4}G_{\delta\!s}(s_{1m}+1)
 +\exp\left(-\frac{1}{2\delta\! s^2}\right)\frac{1}{2}G_{\delta\!s}(s_{1m})
 + \frac{1}{4}G_{\delta\!s}(s_{1m}-1)
\nonumber \\ 
P(s_{1m};s_2=-1) &=& \frac{1}{4}G_{\delta\!s}(s_{1m}+1)
 -\exp\left(-\frac{1}{2\delta\! s^2}\right)\frac{1}{2}G_{\delta\!s}(s_{1m})
 + \frac{1}{4}G_{\delta\!s}(s_{1m}-1).
\end{eqnarray}
Each Gaussian contribution to the joint probabilities given in equation 
(\ref{eq:sumcp}) can be identified with elements of the density matrix 
of the original state in the eigenstate basis of the observable $\hat{s}_1$.
As discussed in a previous paper \cite{Hof00a}, the measurement of
$s_{1m}$ modifies each matrix element by a decoherence factor 
given by the difference of the eigenvalues and an information factor 
depending on the difference between the measurement result $s_{1m}$ and 
the average of the eigenvalues. 

The decoherence factor $\exp(-1/(2\delta\!s^2))$ is a result of the
quantum noise in the measurement required by the uncertainty principle. 
In the case of the beam displacer acting on single 
photons, it is the uncertainty of the wave vector dependent time the photon 
spends in the birefringent medium which randomly rotates the Stokes vector 
around the $s_1$ axis. 
Since this noisy interaction is statistically independent of
the measurement result, it is possible to separate its effect 
from the information obtained about the system. A hypothetical noise free 
measurement then reveals negative probabilities of $s_2=-1$ for measurement 
results $s_{1m}$ close to zero \cite{Hof00a}. These negative probabilities
describe the non-classical correlations between non-commuting operator
variables \cite{Hof00b,Hof00c}. 

The information about $\hat{s}_1$ obtained in the measurement 
modifies the statistical weight of each density matrix element
by a Gaussian function of the difference between the 
measurement result $s_{1m}$ and the average of the two eigenvalues
of the density matrix element. In particular, the Gaussians centered 
around $s_{1m}=0$ represent contributions from
the quantum coherence between the $s_1=+1$ and the $s_1=-1$ 
eigenstates conditioned by a measurement of $s_{1m}$. 
Measurement results close to $s_{1m}=0$ enhance the coherence 
and increase the probability of $s_2=+1$ to values above 1, while 
measurement results far away from $s_{1m}=0$ reduce the coherence, 
lowering the probability of $s_2=+1$ to values below 1. In order to explain
this non-classical correlation between $s_1$ and $s_2$, some measure of 
reality must be attributed to $s_1=0$, even though it is not an 
eigenvalue of $\hat{s}_1$ \cite{Hof00b}. 
Since the width of the Gaussians represents the effect of random noise 
in the readout of the finite resolution measurement, it is reasonable 
to identify each Gaussian contribution with its average value of $s_{1m}$. 
The continuum of measurement values $s_{1m}$ can then be represented by
a discrete set of three values at $s_1=\pm 1$ and $s_1=0$. The joint
probabilities for these three values of $s_1$ and the two eigenvalues
of $s_2$ read
\begin{eqnarray}
\label{eq:joint0}
P(s_1=-1;s_2=-1)=1/4 &\hspace{1cm}& P(s_1=-1;s_2=+1)=1/4
\nonumber \\
P(s_1=0;s_2=-1)=-1/2 &\hspace{1cm}& P(s_1=0;s_2=+1)=1/2
\nonumber \\
P(s_1=+1;s_2=-1)=1/4 &\hspace{1cm}& P(s_1=+1;s_2=+1)=1/4.
\end{eqnarray}
These joint probabilities explain the non-classical features of the
quantum statistics obtained from the single photon polarization 
measurement setup shown in figure \ref{setup1} for any value of the
measurement resolution $\delta\!s$.

It should be noted that the measurement setup itself defines an 
asymmetry between $\hat{s}_1$ and $\hat{s}_2$ 
since the non-eigenvalue of zero appears only in the statistics of 
the initial finite resolution measurement of $\hat{s}_1$. 
This dependence on the order of
measurement is reflected in the operator order dependence of quantum
mechanical expectation values. In order to identify the operator properties 
responsible for the appearance of negative probabilities in the statistical 
properties, it is useful to characterize the measurement
statistics in terms of the correlation between $s_{1m}^2$ and $s_2$, 
\begin{eqnarray}
\label{eq:corr}
C(s_{1m}^2,s_2) &=& \langle s_{1m}^2 s_2 \rangle - \langle s_{1m}^2 \rangle
\langle s_2 \rangle
\nonumber \\[0.2cm]
&=& - 2 (\langle s_{1m}^2 \rangle-\delta\!s^2) \langle s_2 \rangle
\nonumber \\
&=& - 2 \exp\left(- \frac{1}{2\delta\!s^2}\right),
\end{eqnarray}
where $\langle \; \rangle$ denotes statistical averages over 
actual measurement results.
This correlation may be expressed in terms of the operator expectation
values of $\mid \psi_{\mbox{in}}\rangle$ as
\begin{equation}
C(s_1^2,s_2) = \exp \left( - \frac{1}{2\delta\!s^2}\right)
\left(\langle \psi_{\mbox{in}}\mid \hat{s}_1\hat{s}_2\hat{s}_1
\mid \psi_{\mbox{in}} \rangle - 
\langle \psi_{\mbox{in}} \mid \hat{s}_1^2 \mid \psi_{\mbox{in}} \rangle 
\langle\psi_{\mbox{in}} \mid \hat{s}_2 \mid \psi_{\mbox{in}}\rangle \right). 
\end{equation}
As explained above, the exponential factor expresses the randomization of 
$\hat{s}_2$ induced by the measurement of $\hat{s}_1$ according to the 
uncertainty principle. For $\delta\!s \to \infty$, the noise introduced in
the measurement of $\hat{s}_1$ goes to zero and the correlation is
given by the operator expectation values of the initial state. Due to the 
operator ordering, the anti-correlation between $s_1^2$ and $s_2$ is 
an inherent statistical property of $\mid \psi_{\mbox{in}}\rangle$
even though $\mid \psi_{\mbox{in}}\rangle$ is an eigenstate
of $\hat{s}_2$. 
Thus operator ordering allows a correlation between
fluctuating properties and seemingly well defined operator variables of 
the quantum state. 

This property implies that even the eigenvalues of a
quantum state do not represent ``elements of reality''. Consequently, 
it is wrong to assign measurement values to physical properties before 
the measurement {\it even if the measurement result can be predicted 
with certainty}. Since the violation of Bell's inequalities depends 
on the assignment of such ``elements of reality'' it is not surprising
that it can be violated by quantum theory. 
In the following, it will be shown how the violation of Bell's inequalities 
can be explained in terms of negative joint probabilities obtained from
finite resolution measurements.

\section{Measurement of polarization entanglement}
\label{sec:entangle}
\subsection{Entangled photons}
Entangled photon pairs can be created in two photon emission 
\cite{Fre72,Asp82} or in parametric down conversion \cite{Ou88,Shi88}. 
The precise polarization statistics
may vary depending on the geometry of the setup. In order to express
the violation of Bell's inequalities in terms of the Stokes parameters
$\hat{s}_1$ and $\hat{s}_2$, it is useful to rotate the polarizations
of the two photons in such a way that the quantum state is given by
\begin{equation}
\label{eq:eprstate}
\mid \psi_{a,b}\rangle = \frac{1}{\sqrt{2}}\left(\mid R;L \rangle
+ \exp\left(-i \frac{\pi}{4}\right) \mid L;R \rangle \right).
\end{equation}
This state is an eigenstate of two polarization correlations,
\begin{eqnarray}
\frac{1}{\sqrt{2}}\left(\hat{s}_1(a)+\hat{s}_2(a)\right)\;\hat{s}_1(b)
\mid \psi_{a,b}\rangle &=& \mid \psi_{a,b}\rangle
\nonumber \\
- \frac{1}{\sqrt{2}}\left(\hat{s}_1(a)-\hat{s}_2(a)\right)\;\hat{s}_2(b)
\mid \psi_{a,b}\rangle &=& \mid \psi_{a,b}\rangle.
\end{eqnarray}
The sum of these two eigenvalues violates a Bell's inequality of the form
\begin{equation}
\label{ineq:bell}
K = s_1(a)s_1(b)+s_2(a)s_1(b)-s_1(a)s_2(b)+s_2(a)s_2(b) \leq 2.
\end{equation}
It is therefore not possible to interpret the polarization statistics by
assigning eigenvalues of $\pm 1$ to each Stokes parameter. However, as
indicated by the results of finite resolution measurements on the 
polarization of single photons discussed in section \ref{sec:postulate}
above, such an identification of physical properties with their eigenvalues
is not even consistent with the correlated statistics of local single
photon properties.
The non-classical statistical properties responsible for the violation
of Bell's inequality (\ref{ineq:bell}) can be derived in detail 
by applying the finite resolution measurement setup introduced above
to realize a polarization measurement on the entangled photon pairs
given by equation (\ref{eq:eprstate}). 

\subsection{Experimental setup and measurement statistics}
Figure \ref{setup2} shows the experimental setup for a measurement of the 
correlations in the Bell's inequality (\ref{ineq:bell}). The detector arrays
record coincidence counts between the right and left hand side. Each detector
array corresponds to an eigenvalue measurement of $\hat{s}_2$. The spatial
coordinate at which the photon is registered corresponds to the continuous
measurement value $s_{1m}$ of $\hat{s}_1$. Each measurement result can then be 
identified with a point in one of four two dimensional graphs. 
The probability distribution for the measurement outcomes of the joint
measurements may be determined by projections onto the non-orthogonal, 
non-normalized set of states 
\begin{eqnarray}
\lefteqn{
\mid s_{1m}(a); s_{1m}(b);s_2(a)=\pm 1;s_2(b)=\pm 1
\rangle =} \nonumber \\ && 
\hat{P}_{\delta\!s}(s_{1m}(a))\hat{P}_{\delta\!s}(s_{1m}(b))
\frac{1}{2}\left(\mid R;R \rangle + s_2(a) i \mid L;R \rangle
+ s_2(b) i \mid R;L \rangle - s_2(a)s_2(b)  \mid L;L \rangle
\right).
\end{eqnarray}
In their most compact form,
the joint probabilities of the measurement results $s_{1m}(a)$,
$s_{1m}(b)$, $s_2(a)$, and $s_2(b)$ for the entangled input state 
$\mid \psi(a,b)\rangle$ given by equation (\ref{eq:eprstate}) 
read
\begin{eqnarray}
\label{eq:totprop}
\lefteqn{P(s_{1m}(a); s_{1m}(b);s_2(a)=\pm 1;s_2(b)=\pm 1)=}
\nonumber \\ &\hspace{1cm}&
\frac{\sqrt{2}}{16\pi \delta\!s^2} 
\exp\left(-\frac{s_{1m}(a)^2+s_{1m}(b)^2+2}{2\delta\!s^2}\right)
\nonumber \\ && \times \Big(
2 \sinh\left(\frac{s_{1m}(a)s_2(b)-s_{1m}(b)s_2(a)}{2 \delta\!s^2}\right)
  \cosh\left(\frac{s_{1m}(a)s_2(b)+s_{1m}(b)s_2(a)}{2 \delta\!s^2}\right)
\nonumber \\&& + 
\left(\sqrt{2}+s_2(a)s_2(b)\right) 
   \cosh^2\left(\frac{s_{1m}(a)s_2(b)+s_{1m}(b)s_2(a)}{2 \delta\!s^2}\right)
\nonumber \\&& + 
\left(\sqrt{2}-s_2(a)s_2(b)\right) 
   \sinh^2\left(\frac{s_{1m}(a)s_2(b)-s_{1m}(b)s_2(a)}{2 \delta\!s^2}\right)
\Big).
\end{eqnarray}
Figure \ref{condprop2} shows the results for a measurement resolution of 
$\delta\!s=0.6$. At this intermediate resolution, quantum mechanical
interference effects are especially visible \cite{Hof00b}. 
In particular, separate peaks can be resolved clearly, but quantum 
interference effects are visible in the asymmetric peak shapes
and in the zero probability valleys in the $s_{1m}\approx 0$ regions
separating the peaks. As in the single photon case 
discussed in section \ref{sec:postulate} above, it is indeed possible 
to interpret these features entirely in terms of Gaussian distributions. 
However, negative probability contributions centered around values of 
$s_{1m}=0$ have to be included in order to explain the asymmetries and 
the regions of extremely low probabilities near $s_{1m}=0$ separating
the peaks corresponding to quantized results around $s_{1m}=\pm1$.

\subsection{Violation of Bell's inequality by the finite resolution 
measurement statistics}
As in the one photon case, the regions of low probability at values of 
$s_{1m}(a/b)=0$ can be traced
back to negative joint probabilities. The measurement probabilities
given by equation (\ref{eq:totprop}) may be expressed as a sum of shifted 
normalized Gaussian distributions 
\begin{equation}
G_{\delta\!s}(s_{1m}(a)-d_a;s_{1m}(b)-d_b) := 
\left(2\pi \delta\!s^2\right)^{-1}
\exp\left(-\frac{(s_{1m}(a)^2-d_a)^2+
        (s_{1m}(b)^2-d_b)^2}{2\delta\! s^2}\right). 
\end{equation}
Since the shifts $d_a$ and $d_b$ may be $-1$, $0$ or $+1$, respectively,
each of the four sums has nine components associated with joint probabilities
of the four Stokes parameters. The probability distribution of the measurement
results is then given by the sum  
\begin{eqnarray}
\label{eq:eprcond}
\lefteqn{P(s_{1m}(a);s_{1m}(b);s_2(a)=\pm1;s_2(b)=\pm1) 
=}
\nonumber \\[0.2cm] &\hspace{1cm}&
\frac{\sqrt{2}+1}{16\sqrt{2}} 
\left(G_{\delta\!s}\left(s_{1m}(a)+1;s_{1m}(b)+1\right)+
 G_{\delta\!s}\left(s_{1m}(a)-1;s_{1m}(b)-1\right)\right)
\nonumber \\ && +
\frac{\sqrt{2}-1}{16\sqrt{2}} 
\left(G_{\delta\!s}\left(s_{1m}(a)+1;s_{1m}(b)-1\right)+
 G_{\delta\!s}\left(s_{1m}(a)-1;s_{1m}(b)+1\right)\right)
\nonumber \\ && + \exp\left(-\frac{1}{2\delta\!s^2}\right) 
\frac{1}{8\sqrt{2}} s_2(b) 
\left(G_{\delta\!s}\left(s_{1m}(a)+1;s_{1m}(b)\right) -
G_{\delta\!s}\left(s_{1m}(a)-1;s_{1m}(b)\right)\right) 
\nonumber \\ && - \exp\left(-\frac{1}{2\delta\!s^2}\right) 
\frac{1}{8\sqrt{2}}s_2(a)
\left(G_{\delta\!s}\left(s_{1m}(a);s_{1m}(b)+1\right) -
G_{\delta\!s}\left(s_{1m}(a);s_{1m}(b)-1\right)\right)
\nonumber \\ && + \exp\left(-\frac{1}{\delta\!s^2}\right) 
\frac{1}{4\sqrt{2}} s_2(a) s_2(b)\; 
G_{\delta\!s}\left(s_{1m}(a);s_{1m}(b)\right). 
\end{eqnarray}
Using this decomposition, it is a straightforward matter to determine the
averages corresponding to the correlations of the Bell's inequality 
(\ref{ineq:bell}) by summing over $s_2(a)$ and $s_2(b)$ and integrating over
the continuous results $s_{1m}(a)$ and $s_{1m}(b)$. The result reads
\begin{equation}
\label{eq:kav}
\langle K \rangle =
\frac{1}{\sqrt{2}}\left(1+\exp\left(-\frac{1}{2\delta\!s^2}\right)\right)^2. 
\end{equation}
This expectation value exceeds the maximal value of 2 allowed by inequality 
(\ref{ineq:bell}) for measurement resolutions of $\delta\!s > 1.143$. 
The violation of Bell's inequality can therefore be obtained directly from 
the measurement statistics for sufficiently low resolutions of the 
$\hat{s}_1$ measurements. An example for this direct violation of 
Bell's inequality is shown in figure \ref{lowres} for a measurement 
resolution of $\delta\! s=2$. At this low resolution, quantization effects
are not resolved. The non-classical properties of the statistics are
observable in the shift of the peak maxima for $s_2(a)= - s_2(b)$ to values 
greater than $s_{1m}=+1$ or lower than $s_{1m}=-1$. Specifically, the 
maximum probability density for $s_2(a)=+1$ and $s_2(b)=-1$ is at 
$s_{1m}(a)==s_{1m}(a)=1.383$ and the maximum for $s_2(a)=-1$ and 
$s_2(b)=+1$ is at $s_{1m}(a)==s_{1m}(a)=-1.383$. The value of $K$
at these points would be equal to 3.68. 

While it might be tempting to interpret the statistics in terms of 
polarization components greater than $+1$ or smaller than $-1$, 
the high resolution results of figure \ref{condprop2} and the
analysis of single photon polarization in section \ref{sec:postulate}
suggests that the true reason for the shifted peaks are negative
probabilities around $s_{1m}(a)=s_{1m}(b)=0$. In order to obtain
a consistent interpretation of the measurement results for both
high and low resolutions, it is necessary to identify the decoherence
factor $\exp(1/(2\delta\!s^2))$ with the quantum noise induced 
reduction of the expectation values of $\hat{s}_2(a)$ and $\hat{s}_2(b)$.
It is then possible to remove the effects of noise and of finite measurement
resolution from the measurement statistics, tracing the violation of 
Bell's inequality directly to the appearance of negative probabilities in the 
joint probabilities for $s_1(a)$,$s_2(a)$,$s_1(b)$, and $s_2(b)$.

\section{Disentangling entanglement: interpretation of the non-classical
statistics}
\label{sec:quantcorr}
\subsection{Negative conditional probabilities in photon entanglement}
As in the case of single photon polarization discussed in section 
\ref{sec:postulate}, the sum of Gaussians given in equation 
(\ref{eq:eprcond}) can be interpreted in terms of joint probabilities 
for $s_1(a)$,$s_2(a)$,$s_1(b)$, and $s_2(b)$ by 
identifying the average of each Gaussian with the appropriate value of
$s_1$. The joint probabilities for all 36 combinations of the six
contributions from $s_1(a)$ and $s_2(a)$ with the six contributions from
$s_1(b)$ and $s_2(b)$ characterizing the statistics of the 
entangled state $\mid \psi_{a,b}\rangle$ are shown in table \ref{negprop}.
From these probabilities, the statistical weight of different
contributions to the sum correlation $K$ in inequality (\ref{ineq:bell})
can be determined.

The joint probabilities can be classified according to whether the values
of $s_1(a)$ and $s_1(b)$ are zero or not. There are sixteen contributions with
both $s_1(a)$ and $s_1(b)$ non-zero. These cases correspond to the classical
expectation that the values of $s_1$ should be equal to the eigenvalues
observed in high resolution measurements. Consequently, they are the only
contributions that are not diminished by the decoherence factor for small
$\delta\!s$. Moreover, their probabilities are all positive. In eight of 
these sixteen cases, three of the four correlations in inequality 
(\ref{ineq:bell}) are equal to $+1$ and one is equal to $-1$,
for a total of $K=2$. The reverse is true for the remaining eight cases,
resulting in a total of $K=-2$ for the sum of correlations in inequality
(\ref{ineq:bell}). The probability of each case is equal to 
$(2+s_1(a)s_1(b)\sqrt{2})/32$. Summing up the probability of the eight
cases with $K=2$ thus results in a total probability of $(4+\sqrt{2})/8$ or
roughly 67.7 \%. The eight cases with $K=-2$ have a total probability
of $(4-\sqrt{2})/8$, or 32.3 \%. The average value of $K$ for these 
``classical'' contributions to the joint probability is therefore equal 
to $1/\sqrt{2}$, as evidenced by the limit of equation (\ref{eq:kav}) 
for $\delta\!s \to 0$. Obviously, the violation of Bell's inequality must 
originate from the remaining twenty contributions with at least one value 
of $s_1$ equal to zero. 

There are sixteen contributions with one value of $s_1$ equal to zero and
the other value non-zero. Two of the four correlations in the inequality
(\ref{ineq:bell}) are then equal to zero, while the other two may be either
plus or minus one each. In four cases, they are both equal to plus one ($K=2$),
in eight cases, they have opposite sign ($K=0$), and in the remaining four 
cases, they are both equal to minus one ($K=-2$). The probabilities for these
cases are $\pm \sqrt{2}/16$. As a result, the total probability for the four 
cases with $K=2$ is equal to $\sqrt{2}/4$ or 35.4\%, the total probabilities 
for the eight cases with $K=0$ cancel to zero, and the total probability for
$K=-2$ is $-\sqrt{2}/4$ or -35.4\%. This negative probability more than 
outweighs the 32.3 \% of the classical contributions, explaining the increase
of the expectation value of $K$ beyond the limit of 2. However, the 
effect is further enhanced by the contributions from $s_1(a)=s_1(b)=0$.

There are four contributions with $s_1(a)=s_1(b)=0$. Only the correlation
$\langle s_2(a)s_2(b)\rangle$ is non-zero in these cases. Two cases
have $K=1$ and a positive probability of $\sqrt{2}/8$, and two cases have
$K=-1$ and a negative probability of $-\sqrt{2}/8$. This adds a total 
probability of 35.4 \% for $K=1$ and -35.4 \% for $K=-1$. 

The probability distribution over values of $K$ can be summarized as 
follows:
\begin{eqnarray}
P(K=2) = 103.1 \% &\hspace{1cm}& P(K=-2) = -3.1 \%
\nonumber \\
P(K=1) = 35.4 \% && P(K=-1) = -35.4 \%
\nonumber \\
P(K=0) = 0 \%. &&
\end{eqnarray}
The high expectation value of $K$ is a result of the negative probabilities
for combinations of  $s_1(a)$,$s_2(a)$,$s_1(b)$, and $s_2(b)$ with $K<0$. 
In the measured probability distributions described by equations 
(\ref{eq:totprop}) and (\ref{eq:eprcond}), these negative probabilities appear
as a suppression of the probability for values of $s_1$ close to zero, 
pushing the peak of the probability distribution beyond the eigenvalue 
limit of $\pm 1$. Since $s_{1m}$ is not restricted to eigenvalues of 
$\hat{s}_1$, the contributions to the expectation value of $K$ taken from 
the measured distribution (\ref{eq:totprop}) shown in figure \ref{condprop2} 
may exceed the classical limit. 
Even though a direct observation of the negative 
probabilities is of course impossible, the continuous distribution of finite
resolution measurement results thus reveals clear evidence of these 
non-classical statistical features.

\subsection{Quantum noise and negative probabilities in entangled systems}

The negative conditional probabilities shown in table \ref{negprop}
allow an interpretation of the measurement statistics in terms of 
individual measurement results observed separately in branch $a$ and
in branch $b$. There is neither a need for action at a distance, nor
for non-local properties. The non-classical feature required to 
explain the violation of Bell's inequalities is expressed in the 
negative probabilities which are possible even in individual quantum
systems because the uncertainty principle does not allow an isolated 
measurement of a joint probability of non-commuting variables. 

Once the relationship between uncertainty and negative conditional
probabilities is understood, the problem of non-locality in
entangled systems can be resolved by introducing local decompositions
of the entangled state density matrix based on negative probability 
components of the local density matrices. For the state discussed 
above, one possible decomposition reads
\begin{eqnarray}
\mid \psi_{a,b} \rangle \langle \psi_{a,b} \mid &=& 
\frac{1}{4} \hat{1}(a) \otimes \hat{1}(b)
+ \frac{1}{4\sqrt{2}} (\hat{s}_1(a) + \hat{s}_2(a))\otimes \hat{s}_1(b)
\nonumber \\ &&
+ \frac{1}{4\sqrt{2}} (\hat{s}_1(a) - \hat{s}_2(a))\otimes \hat{s}_2(b)
- \frac{1}{4} \hat{s}_3(a) \otimes \hat{s}_3(b).
\end{eqnarray}
All by themselves, the Stokes parameter operators $\hat{s}_i$ would not
qualify as density matrices because of their negative eigenvalues.
Once negative eigenvalue components are permitted, however, the 
decomposition given above can be interpreted as a separation of
the entangled density matrix into products of local density matrices.
The reason why density matrices with negative probability eigenvalues 
may be used in the decomposition of entangled states is that any
measurement performed on system $a$ mixes contributions to the density
matrix of system $b$ in such a way that the information required to 
isolate the negative conditional probabilities represented by the 
negative eigenvalues is lost. 

Effectively, the uncertainty in system
$a$ necessarily ``covers up'' the negative eigenvalues of the density
matrix components of system $b$ by mixing them with positive components.
Entanglement can therefore be explained by the local properties of 
quantum measurements described previously \cite{Hof00a}. In the light 
of this result, it is not surprising that some applications of quantum
mechanics such as quantum computation can work without entanglement 
\cite{Fit00,Bra99}. The most fundamental property of quantum mechanics 
is not entanglement, but local non-classical correlations represented by 
the operator-ordering dependence of expectation values and the negative
conditional probabilities obtained from finite resolution measurements.

\section{Conclusions}
\label{sec:concl}
The interpretation of quantum statistics cannot be based on the 
assumption that potential measurement results represent 
``elements of reality'' whether the actual measurement is performed
or not. This is not only true for entangled systems, but also for 
combinations of finite resolution measurements performed to obtain the
correlations between non-commuting operator variables in a single
quantum system. As a result, concepts such as photon polarization have 
to be reviewed critically in order to understand the relationship between 
eigenvalues and operator variables. 

The experimental approach proposed 
above allows a direct determination of the non-classical features of the 
polarization statistics for both single photons and entangled pairs. 
Its application to the violation of Bell's inequalities reveals details
of the statistical relationships between all four polarization components.
A full set of conditional probabilities can be obtained from the statistics
of a single measurement, revealing the negative conditional probabilities
that are responsible for the violation of Bell's inequalities. A comparison
with the single photon polarization statistics reveals that such negative
probabilities are also observable in the polarization of a single photon. 
The property responsible for the violation of Bell's inequalities is 
therefore a local feature of quantum statistics. Once the implications of
the operator formalism are accepted, entanglement can be understood as a
special case of the non-classical features observable in local correlations.

\section*{Acknowledgment}
The author would like to acknowledge support from the Japanese Society 
for the Promotion of Science, JSPS, and thank Dr. Takao Fuji for 
helpful discussions of the experimental aspects.


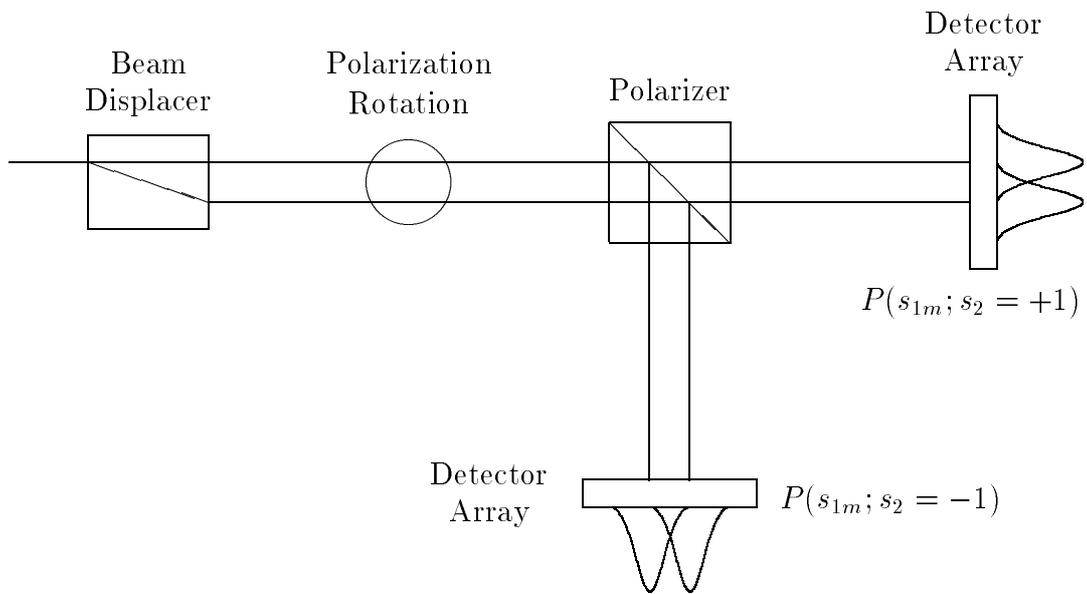
\begin{figure}
\caption{\label{setup1} Schematic representation of the experimental 
setup for a joint measurement of non-orthogonal polarizations. 
The beam displacer separates the incoming light into two parallel beams.
The polarization is then rotated by an angle of $\pi/4$ before the
light beam is split at the polarizer. The overlapping transversal profile 
of the beams is illustrated at the detector arrays.}
\end{figure}
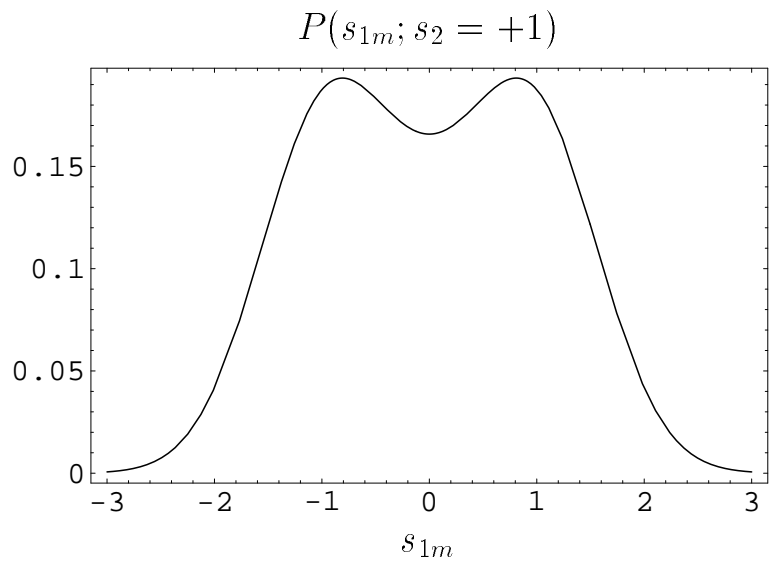
\begin{figure}
\caption{\label{condprop1} Probability distribution $P(s_{1m};s_2)$ for
an initial eigenstate of $s_2=+1$ at a resolution of $\delta\!s=0.6$.
Note the asymmetry and the shifted maxima obtained for $s_2=-1$.}
\end{figure}
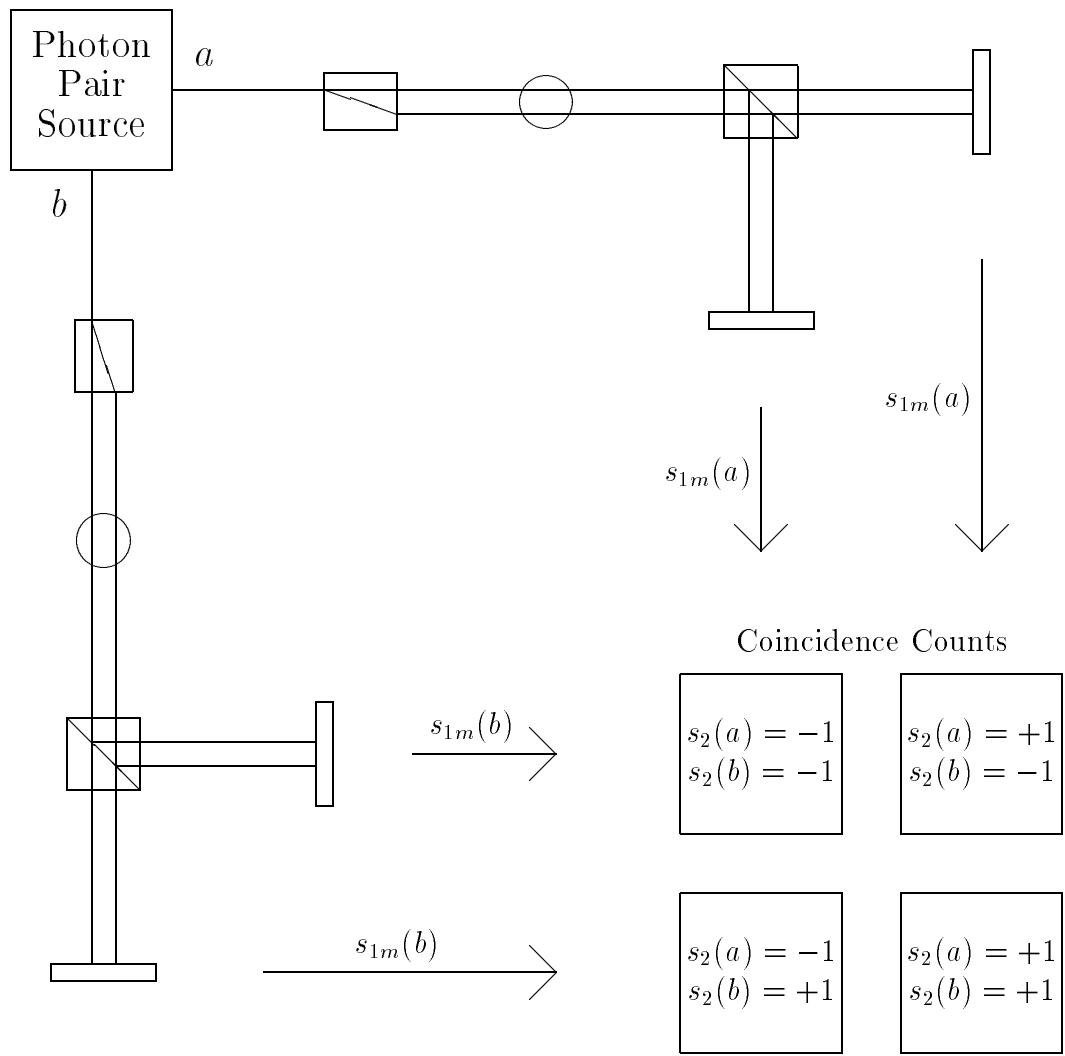
\begin{figure}
\caption{\label{setup2} Schematic representation of the experimental 
setup for a measurement of polarization correlations on entangled photons.
The setup of the branches $a$ and $b$ are as shown in figure 1. Coincidence
counts are registered in one of four channels as illustrated.
}
\end{figure}
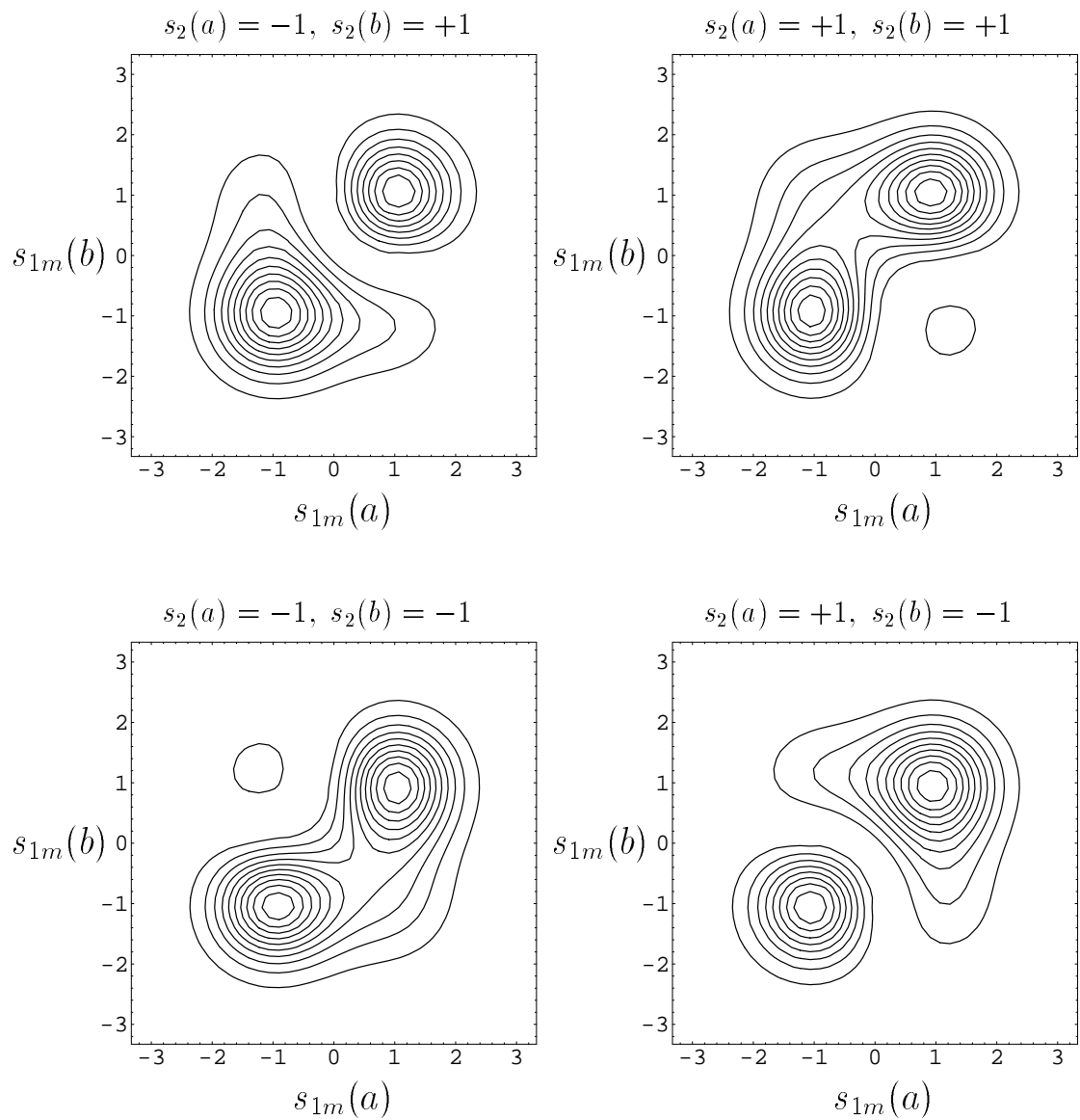
\begin{figure}
\caption{\label{condprop2} Contour plot of the probability distribution 
$P(s_{1m}(a);s_{1m}(b);s_2(a);s_2(b))$ at a resolution of $\delta\!s=0.6$.
While the major peaks appear to be close to the eigenvalues at 
$s_{1m}=\pm1$, the shape of the peaks and the separation between them
reveals the same non-classical statistical effects seen in figure 2.
}
\end{figure}
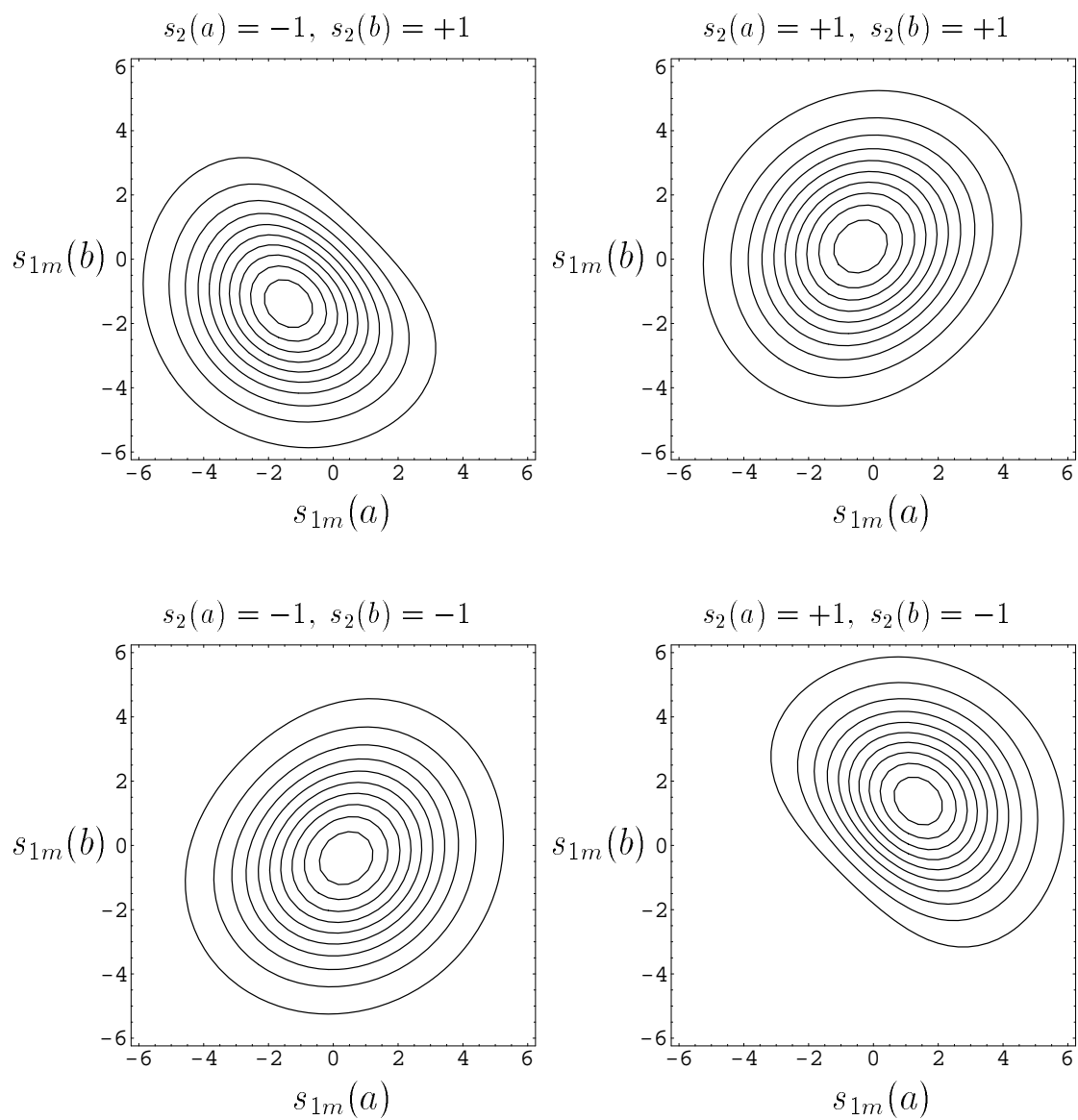
\begin{figure}
\caption{\label{lowres} Contour plot of the probability distribution 
$P(s_{1m}(a);s_{1m}(b);s_2(a);s_2(b))$ at a resolution of $\delta\!s=2$.
The peaks for $s_2(a)=-s_2(b)$ are at $s_{1m}(a)=s_{1m}(b)=\pm 1.383$.
The contribution to the average value of $K$ at these points is  
3.68.}
\end{figure}
\begin{table}
\caption{\label{negprop} Table of conditional probabilities 
derived from the results shown in figure 4. Note that the negative
probabilities roughly coincide with regions of zero probability in 
the measurement statistics.}
\end{table}
%

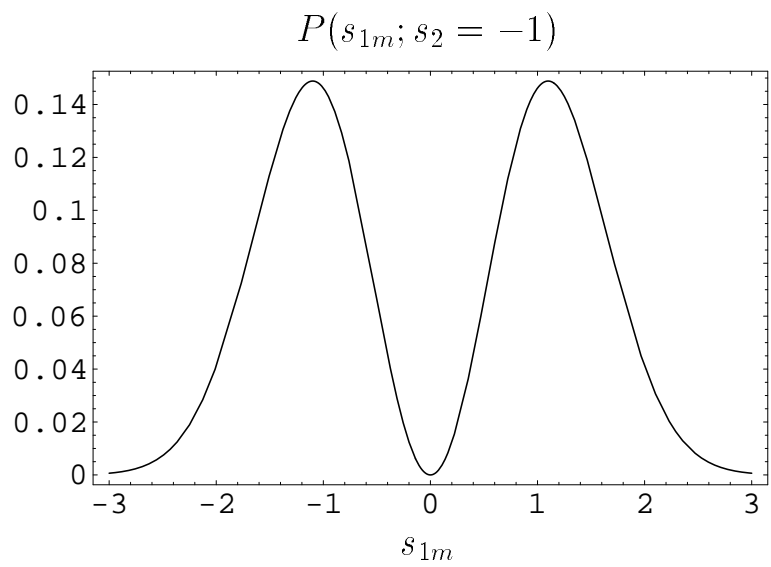
\end{document}